\documentclass[aps,twocolumn,showpacs,pre]{revtex4}
\usepackage{graphicx}
\usepackage{graphics}
\usepackage{amsmath}
\usepackage{amssymb}
\usepackage{array}
\usepackage{longtable}
\usepackage{pstricks}
\usepackage{amsfonts}
\setlongtables
\begin{document}
\title{Spectral Analysis of Protein-Protein Interactions in {\em Drosophila melanogaster}}
\author{Christel Kamp}
\email{mail@christelkamp.de,                           
c.kamp@imperial.ac.uk}
\affiliation{Blackett Laboratory, Imperial College, Prince Consort Road, 
London SW7 2AZ, United Kingdom}
\author{Kim Christensen}
\email{k.christensen@imperial.ac.uk}
\affiliation{Blackett Laboratory, Imperial College, Prince Consort Road, 
London SW7 2AZ, United Kingdom}
\affiliation{Physics of Geological Processes, University of Oslo, PO Box 1048, Blindern, N-0316 Oslo, Norway}

\begin{abstract}
Within a case study on the protein-protein interaction network (PIN) of {\em Drosophila melanogaster} we investigate the relation between the network's spectral properties and its structural features such as the prevalence of specific subgraphs or duplicate nodes  as a result of its evolutionary history.
The discrete part of the spectral density shows fingerprints of the PIN's topological features including a preference for loop structures. Duplicate nodes are another prominent feature of PINs and we discuss their representation in the PIN's spectrum as well as their biological implications.
\end{abstract}
\pacs{89.75.-k, 89.20.-a, 89.75.Hc, 89.75.Fb, 87.16.Yc, 87.16.-b, 87.10.+e, 02.50.Fz}
%more possible pacs
%87.16.Ac, 87.23.-n, 02.60.-x, 02.50.Ng 
\maketitle
\section{Introduction}
Network structures can be observed in most diverse domains ranging from biological and technological systems to social or economical systems \cite{strogatz:2001}. Genetic regulatory networks, protein-protein interaction networks and metabolic networks support the functions of life in any living organism. Technological networks such as the internet or the World Wide Web have a huge impact on our lives and societies. Networks of acquaintances and the exchange of information within these networks shape social and economical systems.
Considering the omnipresence of networks, their
investigation has a long tradition in 
graph theory  \cite{bollobas:modernbook,bollobas:randomgraphs}.
However, during the last few years high quality data on real-world networks 
has revealed that they cannot be adequately described by standard models from random graph theory and the topic has attracted growing interest. 
Still, much attention has been devoted to the derivation of rather 
specific quantities like degree distributions or
clustering coefficients \cite{albert:barabasi:2002} that do not allow for
a classification and understanding of network topologies within a broader and self-consistent framework.
\newline
Making an attempt towards a more comprehensive description 
spectral graph 
theory \cite{chung:book,cvetkovic:book,cvetkovic:annals,biggs:book} 
can be considered as one promising ansatz. 
A network of $N$ nodes 
can be described by its adjacency matrix ${\bf A}=(a_{ij})$ with
entries
\begin{equation}
a_{ij}=\left\{
\begin{array}{ll}
1& \mbox{if there is a link between node $i$ and $j$}\\
0 & \mbox{otherwise}.
\end{array}\right.
\end{equation}
The adjacency matrix is
a symmetric, non-negative matrix in the case of undirected networks 
and accordingly has real eigenvalues $\{\lambda_j\}$, $j=1,...,N$, being solutions of $\det({\bf A}-\lambda{\bf I})=0$. 
The relation between features of a network and properties of its spectral
density 
\begin{eqnarray}
\rho(\lambda)&=&\frac{1}{N}\sum_{j=1}^N\delta(\lambda-\lambda_j)
\end{eqnarray}
with respect to its adjacency matrix 
is a topic of current research. While dense classical random networks
exhibit a  semi-circular spectral density
of the adjacency matrix \cite{bauer:golinelli:2001},
networks with a broad or scale free degree distribution give rise to 
a broader spectrum \cite{farkas:vicsek:2001,dorogovtsev:samukhin:2003,chung:vu:2002,chung:vu:2003,mihail:papadimitriou:2002,goh:kim:2001}.
A striking feature of sparse random networks' spectral density
is the emergence of peaks at eigenvalues of finite 
trees \cite{bauer:golinelli:2001}, similar to those found in 
large random trees \cite{golinelli:2003}, due to the strong prevalence
of these subgraphs \footnote{A similar phenomenon has earlier been discussed in the context of a model of quantum percolation \cite{kirkpatrick:eggarter:1972,evangelou:1983}.}.
Here we address whether these findings are applicable more generally, that is
whether peaks in the spectral density of sparse random networks can be 
associated to a strong prevalence of specific subgraphs.
The search for subgraphs that are statistically
overrepresented relative to a null-model, so-called motifs, 
recently gained much attention \cite{milo:alon:2002,milo:alon:2004}.
As a case study on the relation between these two approaches,
we investigate the spectral properties of the protein-protein interaction 
network (PIN) of the fruit fly {\em Drosophila melanogaster}
\cite{giot:rothberg:2003}. 
While no simple correspondence between network motifs and 
a network's spectral proprieties can be derived, on a more abstract level, 
we infer from the PIN's spectrum a prevalence of loop structures.
Furthermore, some properties specific to
a network that has evolved by duplication of nodes are studied and discussed
within the context of spectral analysis.
%%%%%%%%%%%%%%%%%%%%%%%%%%%%%%%%%%%%%%%%%%%%%%%%%%%%%%%%%%%%%%%%%%%%%%%%%%%%%%
\section{The spectrum of the PIN of {\em Drosophila melanogaster}}
For our study we used the PIN 
of {\em Drosophila melanogaster} as given in \cite{giot:rothberg:2003}
and available via the Database of Interacting Proteins \cite{dip}.  
\begin{figure}[!h]
\begin{pspicture}(0,0)(5,6)%\showgrid
\rput(2.75,3.2){\scalebox{0.33}{\includegraphics{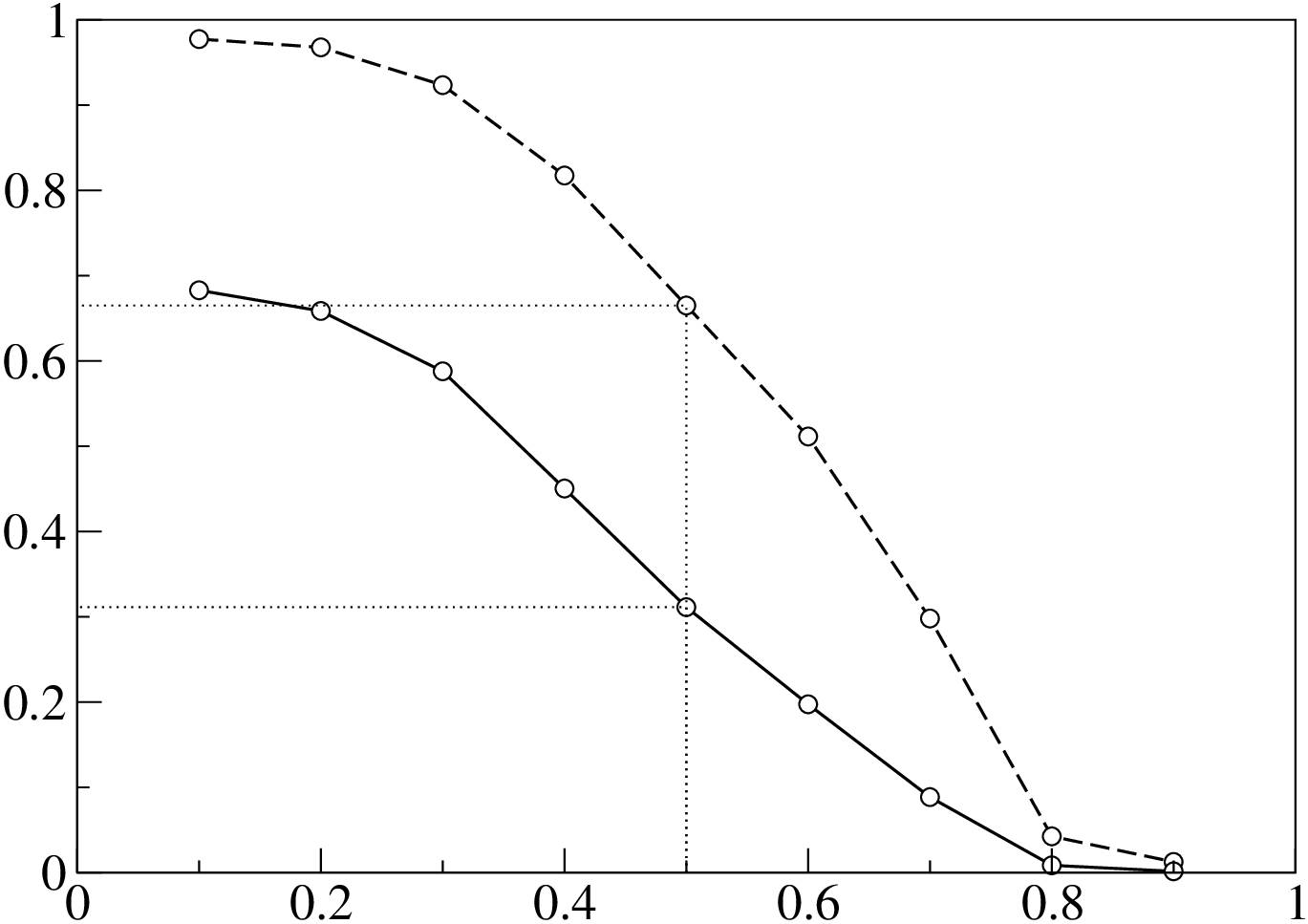}}}
\fontsize{12pt}{12pt}
\rput(3,0.0){Minimal confidence value of interaction}
\end{pspicture}
\caption{The number of nodes divided by 10000 (solid line) and the
fraction of nodes (dashed line) in the largest connected component in the
PIN as a function of the minimal confidence value of protein-protein
interaction. We focus on the PIN defined by a minimal confidence value of $0.5$, see dashed line.
} \label{F:Fig1}
%\end{figure}
%\begin{figure}[!b]
\begin{pspicture}(0,0)(5,12.5)
\rput(2.75,9.5){\scalebox{0.33}{\includegraphics{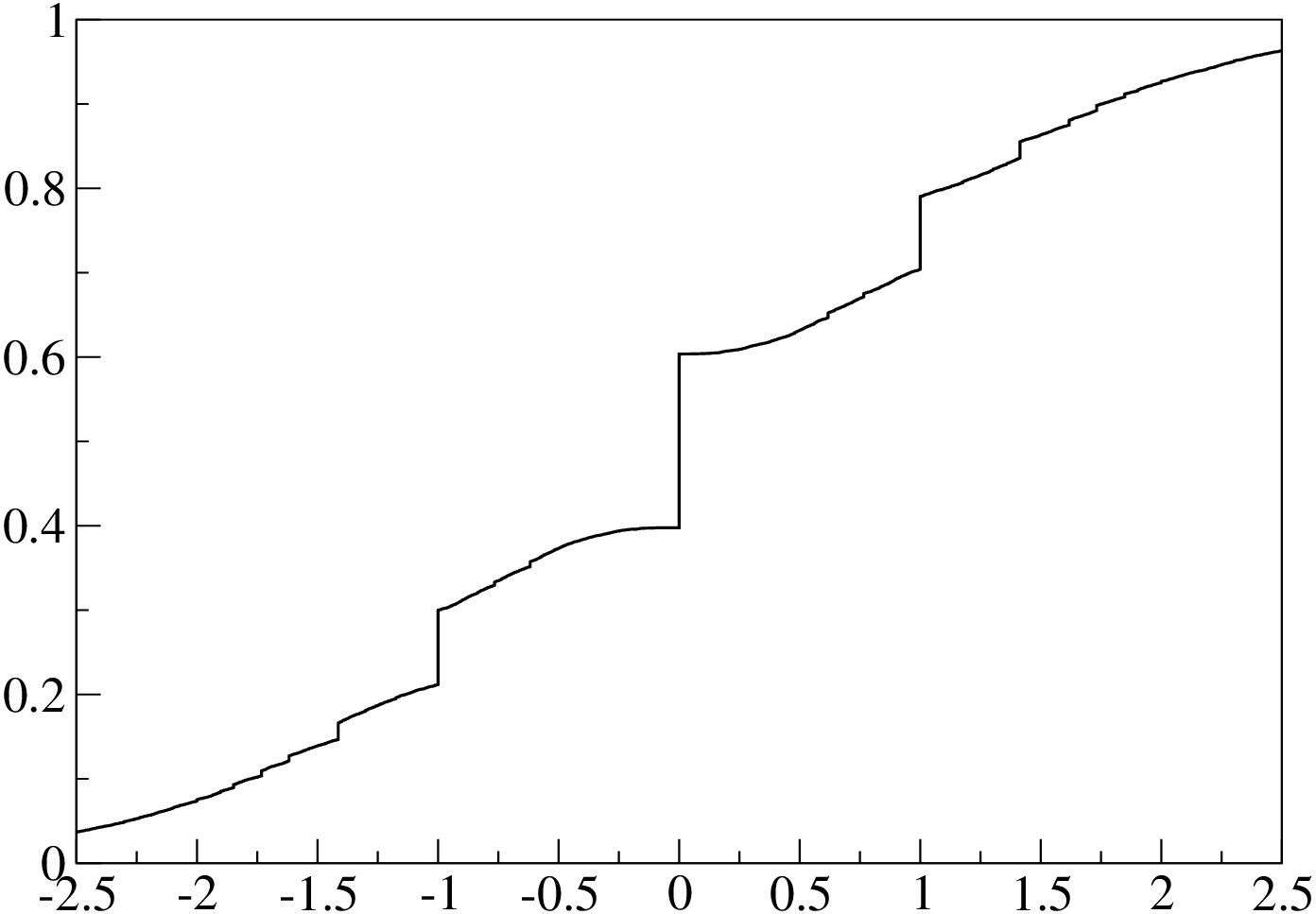}}}
\rput(2.75,3.5){\scalebox{0.33}{\includegraphics{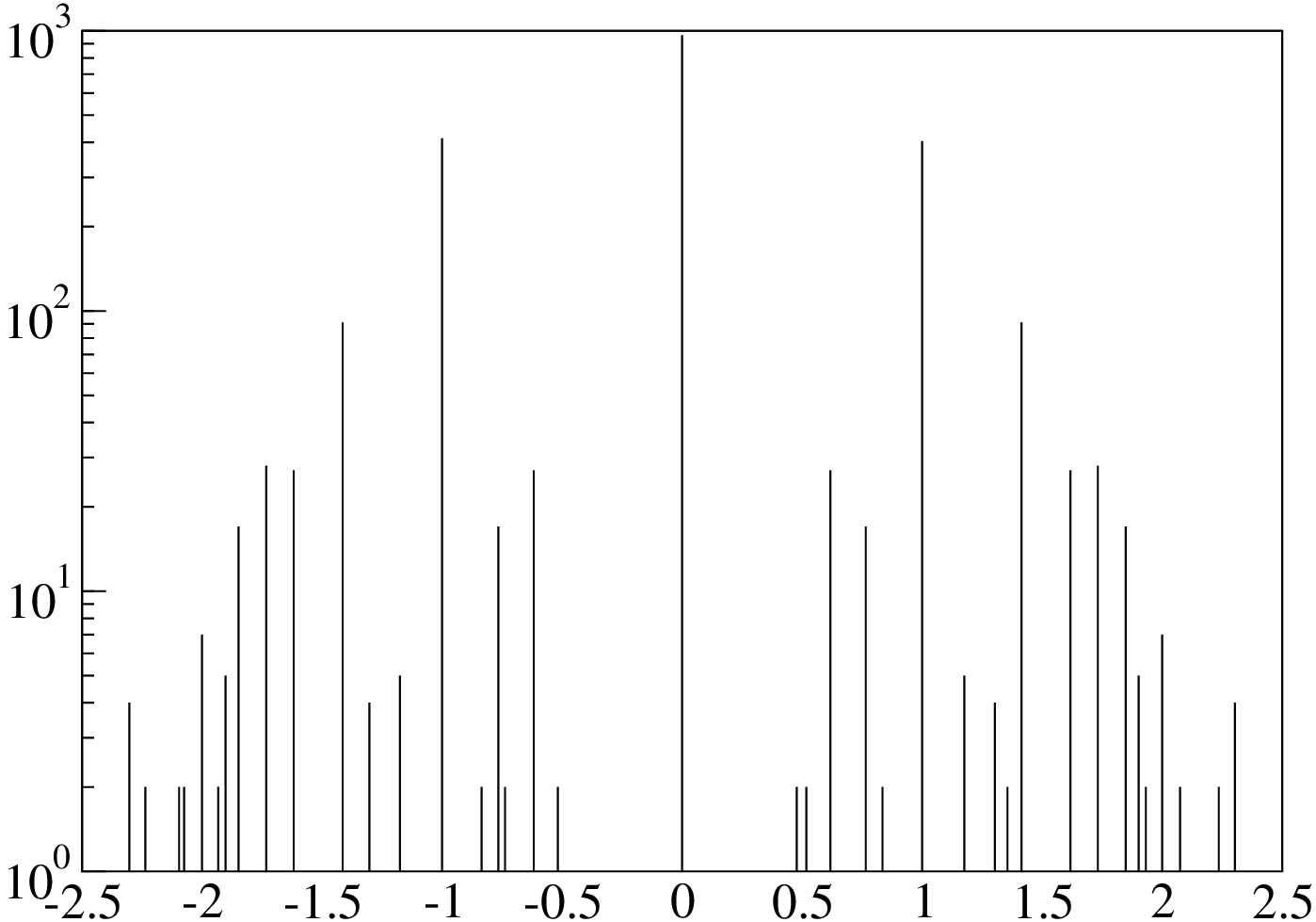}}}
\fontsize{12pt}{12pt}
\rput(2.75,0.25){Eigenvalue}
\rput{90}(-1.65,9.5){Cumulative distribution of eigenvalues}
\rput{90}(-1.65,3.5){Frequency of eigenvalue}
\rput(-0.1,11.6){(a)}
\rput(-0.1,5.6){(b)}
\end{pspicture}
\caption{Spectral analysis of the protein-protein interaction
network of {\em Drosophila melanogaster}. (a) The cumulative
spectral density. (b) The discrete
frequency spectrum containing 49\% of all eigenvalues.  
} \label{F:Fig2}
\end{figure}
The protein-protein interactions have been derived using the 
two-hybrid method which, however, is known to generate many false positives. 
Therefore each interaction in the network is classified by a confidence 
value between zero and one defining a hierarchy of networks with
increasing minimal confidence value for the protein-protein interactions.
In Fig. \ref{F:Fig1} the size of the largest connected component in a network 
with a given minimal confidence value of interactions is shown. 
\newline
For our further analysis we choose a network with a minimal confidence
value of $0.5$ which contains $4681$ proteins and $4794$ interactions 
corresponding to an average degree $\langle k\rangle=2.05$.  
The network is enriched with biologically meaningful 
interactions while it still shows a strong largest connected component (i.e. a giant component) containing 
about $2/3$ of its nodes.
\newline
We determined the eigenvalues of the adjacency matrix corresponding to 
this PIN. The cumulative
spectral density in Fig. \ref{F:Fig2} (a) exhibits
jumps at various eigenvalues which are represented by the discrete
spectrum in Fig. \ref{F:Fig2} (b) \footnote{An eigenvalue is considered
to belong to the discrete spectrum if there exists at least one 
other eigenvalue that does not deviate more than $10^{-12}$.}.
Since about $2/3$ of the network's nodes belong to its
giant component and $49\%$ of the eigenvalues in the network's spectrum
are in the discrete spectrum, the emergence of spectral peaks cannot
be explained by small isolated clusters alone.
%%%%%%%%%%%%%%%%%%%%%%%%%%%%%%%%%%%%%%%%%%%%%%%%%%%%%%%%%%%%%%%%%%%%%%%%%%%%%
\subsection{The discrete spectrum and network motifs}\label{motifsec}
\begin{figure}[!h]
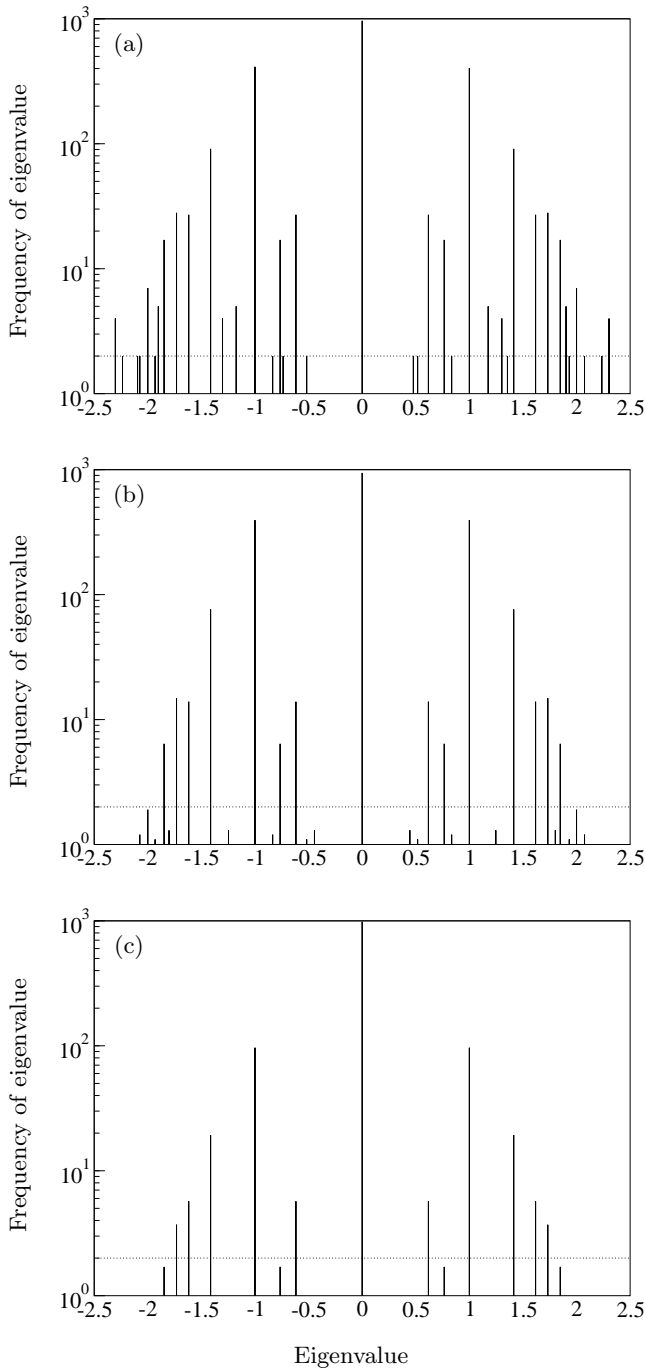

\begin{pspicture}(0,0)(5,18.5)%\showgrid
\rput(2.75,15.5){\scalebox{0.33}{\includegraphics{./Fig3a.eps}}}
\rput(2.75,9.5){\scalebox{0.33}{\includegraphics{./Fig3b.eps}}}
\rput(2.75,3.5){\scalebox{0.33}{\includegraphics{./Fig3c.eps}}}
\fontsize{12pt}{12pt}
\rput(2.75,0.25){Eigenvalue}
\rput{90}(-1.65,15.5){Frequency of eigenvalue}
\rput{90}(-1.65,9.5){Frequency of eigenvalue}
\rput{90}(-1.65,3.6){Frequency of eigenvalue}
\rput(-0.2,17.7){(a)}
\rput(-0.2,11.7){(b)}
\rput(-0.2,5.7){(c)}
\end{pspicture}
\caption{The discrete frequency spectrum of (a) the PIN
of {\em Drosophila melanogaster} containing 49\% of all eigenvalues,
(b) a randomized PIN with identical degrees at each node containing 43\% of all eigenvalues,
and (c) a classical random network of identical size and average degree $\langle k \rangle = 2.05$
containing 27\% of all eigenvalues.
} \label{F:Fig3}
\end{figure}

To get a better understanding of the emergence of spectral peaks
we compare the discrete spectrum with the corresponding spectra 
of two reference networks. First, we look at a network of the same size
and degree
sequence but randomized links following the procedure of 
\cite{maslov:sneppen:2002} (a randomized PIN). Second,
we consider a classical random network of the same size
and average degree 
$\langle k\rangle=2.05$ (a random network), that is a 
network with a probability $p=0.000438$ for a link between any two nodes.
In Fig. \ref{F:Fig3} the discrete spectrum of the adjacency matrix of the 
protein-protein interaction network of {\em Drosophila melanogaster} as well
as of the two reference networks are shown, the latter being averages over $10$ reference networks.

\begin{figure}[b]
\centerline{   \includegraphics[width=6.6cm,angle=0]{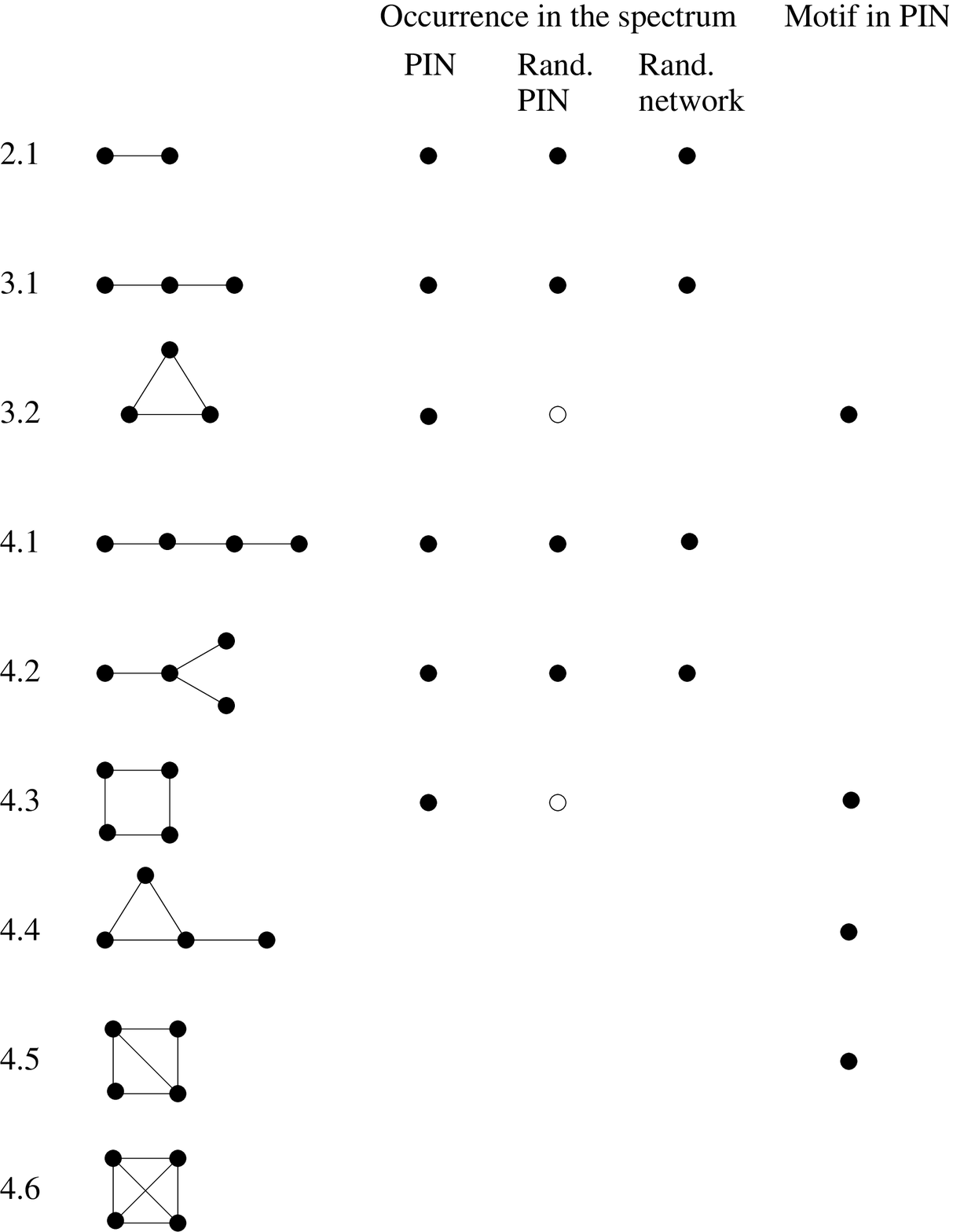}} 
\caption{\label{graphs1} Connected subgraphs with up to 4 nodes: A bullet ($\bullet$) in the three middle columns denotes that the eigenvalues of this graph can be found in the spectrum of the original network (PIN), the randomized network (Rand. PIN) or the random network (Rand. network), respectively. The rightmost column shows whether the subgraph is a motif according to the {\tt mfinder} software (default settings, \cite{milo:alon:2002,mfinder}). White bullets ($\circ$) correspond to single eigenvalue occurrences. }
\end{figure}
\begin{figure}[!h]
\centerline{   \includegraphics[width=6.6cm,angle=0]{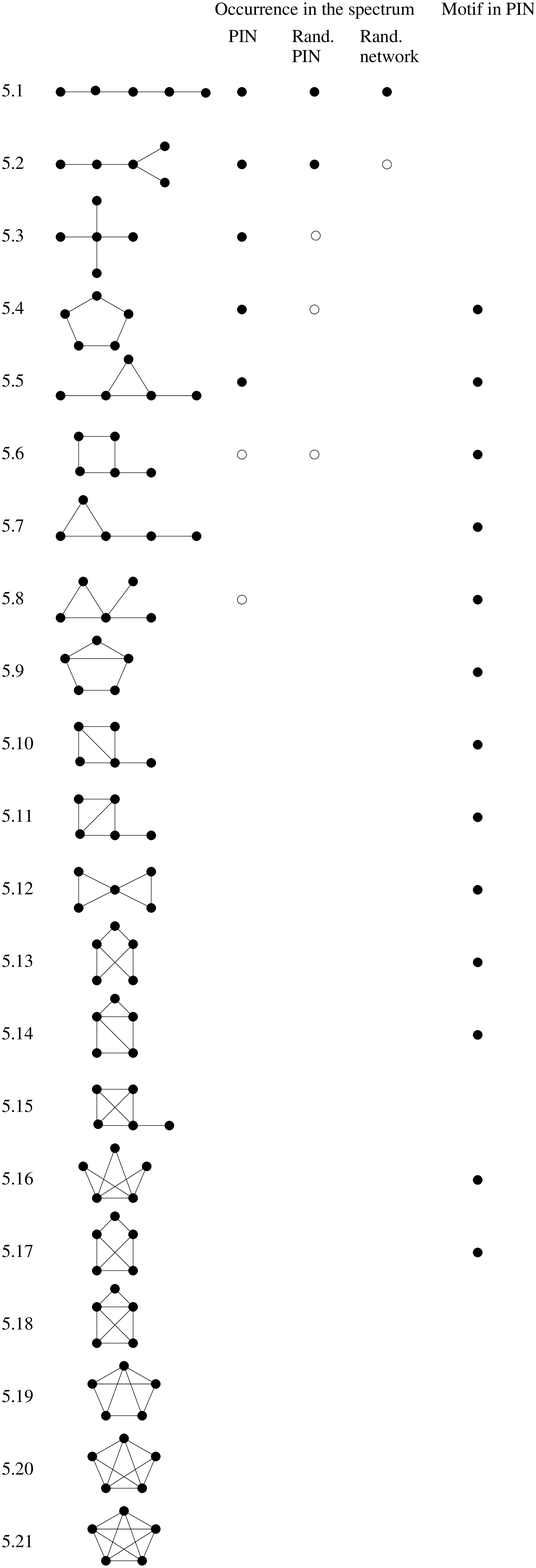}} 
\caption{\label{graphs2}Connected subgraphs with 5 nodes: A bullet ($\bullet$) in the three middle columns denotes that the eigenvalues of this graph can be found in the spectrum of the original network (PIN), the randomized network (Rand. PIN) or the random network (Rand. network), respectively. The rightmost column shows whether the subgraph is a motif according to the {\tt mfinder} software (default settings, \cite{milo:alon:2002,mfinder}). White bullets ($\circ$) correspond to single eigenvalue occurrences. }
\end{figure}

To get more reliable results, we concentrate our further analysis only on 
eigenvalues that can be found more than twice in the spectrum of the original 
network.
Qualitatively, we see that while the classical random network shows only a few peaks of that size
corresponding to the eigenvalues of simple tree-graphs (2.1, 3.1, 4.1, 4.2. in
Fig. \ref{graphs1}), additional eigenvalues appear in the discrete spectrum 
of the randomized PIN  with the same degree sequence as the original network 
and eventually the original network (see Fig. \ref{F:Fig3}). 
This change in the spectral properties indicates some
differences in the structural organization of the underlying networks.
In the following paragraphs, we will discuss how the observed hierarchy of spectral peaks reflects the networks' topologies and relates to other concepts like the search for motifs.
\newline
Following the arguments of \cite{golinelli:2003}, we suggest that 
the prevalence of specific peaks in the discrete spectrum of a network 
corresponds to a strong representation of certain subgraphs.
It has recently been shown that networks from different contexts show
characteristic overrepresentation of  
specific subnetworks which are usually referred to as 
motifs \cite{milo:alon:2002,milo:alon:2004}.
Although motifs can be expected to leave marks in a network's spectrum,
there is seemingly no simple correspondence between the eigenvalues of 
small subgraphs and spectral peaks. First, subgraphs are not generally 
represented by their eigenvalues in the spectrum of the whole network.
Second, isospectral graphs are not necessarily isomorphic 
\cite{spectraofgraphs:book}.
Nevertheless, 
a thorough comparative study of the discrete spectrum can provide some 
insight into the networks' structure. In Figs. \ref{graphs1} and \ref{graphs2} we show the
connected subgraphs up to size $5$ with the full set of eigenvalues present
in the discrete spectrum of the whole network. It shows that the
spectrum of the PIN is more 
consistent with loop-structures (cf. graphs 3.2, 4.3, 5.4, 5.5, 5.6,
5.8) than any randomized version
which might hint to regulatory functionality supplied by this network.
The eigenvalues behind these structures might
correspond to eigenvalues of trees, e.g., the eigenvalues of a triangle
(graph 3.2) or the box (graph 4.3, often also referred to as bi-fan structure)
might well be explained by graphs 2.1 and 5.3. However, to represent
the eigenvalues of graphs 5.5 and 5.6  one has to consider trees of minimum
size 8 and 7, respectively. The eigenvalues of graph 5.8 cannot be found
among trees of size up to 10. Considering that the frequency of a given tree of size $n$ in a sparse network decreases exponentially with $n$ 
\cite{bauer:golinelli:2001}
and relating the findings in the PIN
to those in the randomized reference networks we hypothesize that
the spectral peculiarities reflect the loop structure in the original network.
\newline
To quantify the correspondence between the number of specific subgraphs in the PIN and the PIN's discrete spectrum we tried to decompose the spectrum into the contributions of connected subgraphs up to size $5$. This, however, was not feasible indicating that higher order contributions, though being individually small, cannot be neglected as a whole.
\newline
Although we have to ascertain that there is no simple correspondence between
subgraphs of a network and the prevalence of their eigenvalues in the discrete
spectrum of the whole network we want to discuss the relation of spectral properties to the notion of motifs. According to the definition introduced in Ref. \cite{milo:alon:2002}, a motif is a subnetwork that shows strong prevalence within the network relative to a randomized network. For our analysis we refer to the default requirements implemented in the {\tt mfinder} software \cite{milo:alon:2002,mfinder}, that is a motif is a subgraph that occurs
at least by two standard deviations more than in 100 randomized networks with 
the same degree sequence. 
In Figs. \ref{graphs1} and \ref{graphs2} the rightmost columns show 
which connected subgraphs up to size $5$ are motifs in the PIN according to these criteria.
There exist a lot of highly connected motifs 
while the spectrum reflects more the tree-structures in the network. 
However, the fingerprints of trees in the spectrum of both the 
original and the randomized PIN are consistent with the
fact that they do not show up as motifs according to the above definition.
One might further speculate 
whether some motifs are hidden for spectral analysis because they are
in fact building blocks of larger units. For example graph 5.5 as well as 
its subgraph 4.4 is a motif according to  \cite{milo:alon:2002,mfinder}.
But only the eigenvalues of 5.5 can be found in the spectrum of the PIN.
Moreover highly connected motifs do not occur in high (absolute) numbers
and might accordingly be drowned in spectral analysis.
%%%%%%%%%%%%%%%%%%%%%%%%%%%%%%%%%%%%%%%%%%%%%%%%%%%%%%%%%%%%%%%%%%%%%%%%%%%%%
\subsection{The circuitry of the PIN}
\begin{figure}
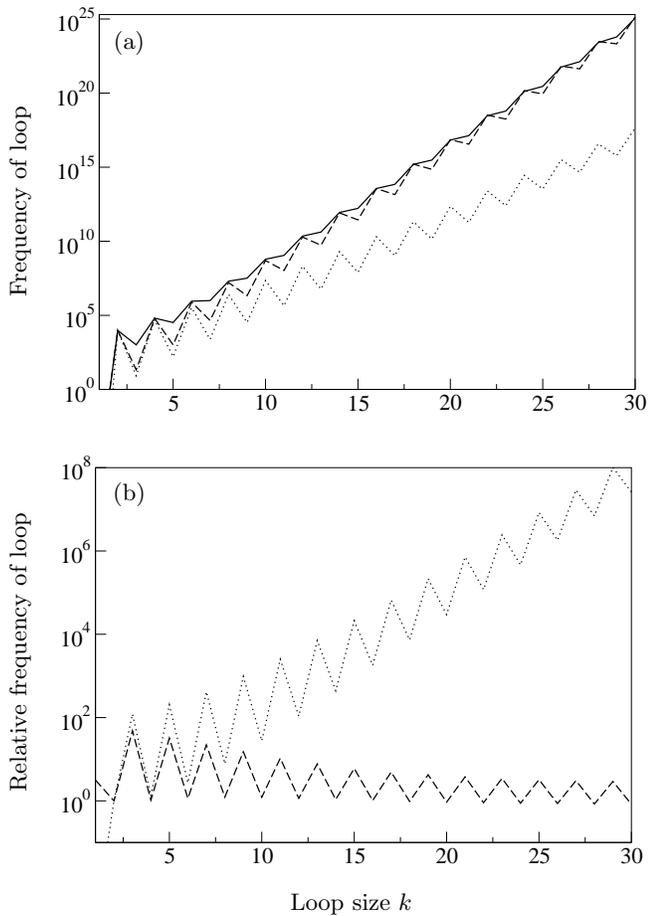

\begin{pspicture}(0,0)(5,12.5)
\rput(2.75,9.5){\scalebox{0.33}{\includegraphics{./Fig6a.eps}}}
\rput(2.75,3.5){\scalebox{0.33}{\includegraphics{./Fig6b.eps}}}
\fontsize{12pt}{12pt}
%\rput(2.75,6.4){Loop size {\small $k$}}
\rput(2.75,0.25){Loop size {\small $k$}}
\rput{90}(-1.65,9.6){Frequency of loop}
\rput{90}(-1.65,3.6){Relative frequency of loop}
\rput(-0.2,11.7){(a)}
\rput(-0.2,5.7){(b)}
\end{pspicture}
\caption{(a) The frequency of loops of size $k$ in the PIN of {\em Drosophila melanogaster} (solid line),
 a randomized PIN with identical degrees at each node (dashed line) and
a classical random network of identical size and average degree $\langle k \rangle = 2.05$ (dotted line).
Odd cycles represents non-trivial loops, that is, deviations from a tree-like structure
in the networks.
(b) The relative frequency of loops of size $k$ in the PIN of {\em Drosophila melanogaster}
with respect to the randomized PIN (dashed line) and the classical random network (dotted line).
} \label{F:Fig6}
\end{figure}

In section \ref{motifsec} we have shown that the discrete spectrum of the PIN of {\em Drosophila melanogaster} favors the eigenvalues of loopy subgraphs.
This observation derived from the investigation of distinct local structures and their eigenvalue representations can be confirmed by an assessment of the whole set of eigenvalues.
Evaluating  the trace of the matrix ${\bf A}^k$
\begin{equation}\label{trace}
Tr({\bf A}^k)=\sum_{i=1}^N \lambda_i^k
\end{equation}
yields the number of directed loops of length $k$ 
in the underlying network \cite{cvetkovic:book,farkas:vicsek:2001} as shown in Fig. \ref{F:Fig6}, though neglecting details of the graphs underlying the loops. Note, that even loops might be trivial going back and forth in a tree while odd loops are non-trivial. 
The difference in growth rates of the numbers of loops of growing size between the original network and the classical random network is likely due to the strong fragmentation of the latter one (many isolated nodes). However, the strong relative prevalence of loops of odd length in the original network is more remarkable with respect to the networks' topologies. This becomes more obvious from 
 Fig. \ref{F:Fig6} (b) showing the number of loops of a given size in the original PIN normalized to the numbers in the two reference networks.  While tree graphs only have trivial loops of even length, loops of odd length indicate non-trivial loops which confirms the results derived from the evaluation of the discrete spectrum on the basis of eigenvalue representations of small subgraphs.
\newline
The analysis of a network's discrete spectrum could reveal some structural information about the network as a whole. This information is less specific than an analysis in terms of motifs, that is only conclusions about more general properties like the prevalence of loops are possible instead of exact motif counts. However, it should be emphasized that spectral analysis is not hampered by an a priori {\em bias} towards predefined
quantities like motifs of a given size.
It is a challenging question of future research to investigate the relationship between a network's spectrum and its topological features, e.g. in terms of motifs, in more detail to get a more rigorous and {\em unbiased} characterization of a network's topological features. 
%%%%%%%%%%%%%%%%%%%%%%%%%%%%%%%%%%%%%%%%%%%%%%%%%%%%%%%%%%%%%%%%%%%%%%%%%%%%%%
\section{Fingerprints of duplication}\label{dup}
The evolution of many biological networks and specifically PINs 
is assumed to be strongly driven by duplication (and diversification)
of nodes in the network \cite{ohno:book,teichmann:babu:2004}.
The genomes underlying the PIN of many organisms have undergone a few whole genome duplications complemented by many single-gene duplications \cite{amoutzias:bauer:2004}. After duplication, one of the duplicates usually diverges from its original appearance, possibly providing new functionality. 
The concept of duplication has similarly been recognized to be important for  
functional roles in a network motif \cite{kashtan:alon:2004}. 
The search for fingerprints of the evolutionary history of a PIN naturally has to include an assessment of duplicate nodes, that is those that share the same interaction partners. Each set of duplicate nodes represents an equivalence class also referred to as an orbit. The reduced network is a network in which all nodes of an orbit are reduced to one node.
\begin{table}[h]
\begin{tabular}{l|l|l}
& Duplicate nodes & Duplicate links\\
\hline
PIN {\em D. melanogaster} & $686$ & $728$\\
Randomized PIN & $626.0\pm 22.0$ & $629.1\pm 22.0$\\
Random network & $151.0\pm14.2$ & $151.0 \pm 14.2$\\
\end{tabular}
\caption{\label{duptab} The table shows the number of nodes
that are duplicates (duplicate nodes) and the 
number of neighbors associated to these nodes (duplicate links) in the original network and the two reference networks, that is the difference in the number of nodes and links between the original network and the reduced network. Isolated nodes have been neglected.}
\end{table}

\begin{figure}[!h]
\begin{pspicture}(0,0)(5,6.5)%\showgrid
\rput(2.75,3.5){\scalebox{0.33}{\includegraphics{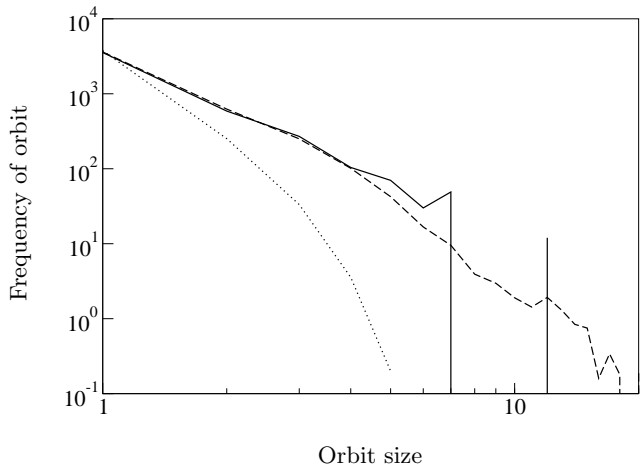}}}
\fontsize{12pt}{12pt}
\rput(3,0.25){Orbit size}
\rput{90}(-1.65,3.6){Frequency of orbit}
\end{pspicture}
\caption{The frequency of orbit size in the PIN of {\em Drosophila melanogaster} (solid line),
a randomized PIN with identical degrees at each node (dashed line) and
a classical random network of identical size and average degree $\langle k \rangle = 2.05$ (dotted line).
Isolated nodes  have been neglected.
} \label{F:Fig8}
\end{figure}

Tab. \ref{duptab}
shows that the PIN has more duplicate
nodes (with associated links) than the reference networks.
Fig. \ref{F:Fig8} shows the 
frequency of orbits of a given size in the original as well as in 
the reference networks. The distribution of orbit sizes in the original
network is very close to the one found in the randomized 
PIN with the same degree sequence, but much broader than 
that of the classical random network.

Again, spectral analysis offers a complementary approach to the topic. 
Using the results in Appendix A, we determined the eigenvalues
of those graphs that arise from duplication of the two simplest reduced
graphs: a line and a triangle (graphs 2.1 and 3.2 in Fig. \ref{graphs1}). 
We allowed for up to ten duplications of each
node of the reduced network and searched for the eigenvalues of the 
resulting subgraphs. However, spectral analysis is only consistent
with the emergence of star graphs and the original triangle as
well as the box or bi-fan structure (graph 4.3 in Fig. \ref{graphs1}).
\newline
Considering the representation of star graphs in the 
spectrum of the PIN one might guess
that the high frequency of large orbits mainly reflects nodes with many leaves.
A look at the joint distribution of the size of an orbit and the degree
of its nodes in the original and the reference networks supports this 
hypothesis. In 100 reference networks (of both kind) 
the nodes in an orbit larger than one have degree one,
that is only nodes with degree one have duplicates.
Only in extremely rare cases do nodes with degree two have a single duplicate.
\begin{figure}[t]
\includegraphics[width=8cm,angle=0]{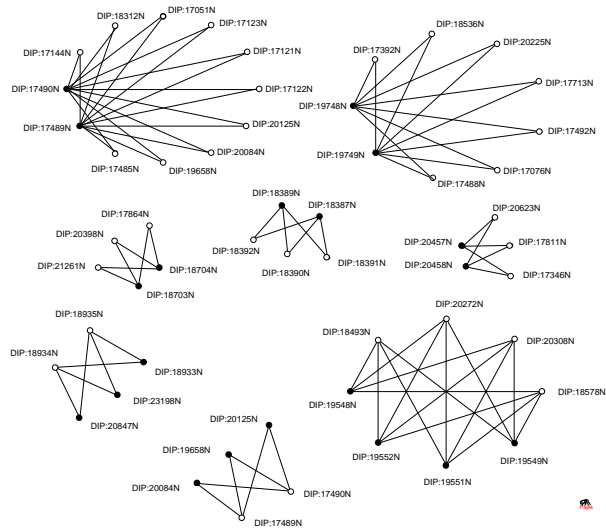}
\caption{\label{dupgraphs}Subgraphs of nodes that form orbits of 
size $\geq 2$ and that have
degree $\geq 2$ (black), orbits of size 2 of nodes with degree 2 have been omitted. The white nodes are the neighbors of duplicate nodes that may have
more neighbors than shown. The nodes' labels are their identifiers from the Database of Interacting Proteins \cite{dip}, cf. Appendix B for more details.}
\end{figure}
\newline
This matches the global situation in the original network, however,  there are 
some remarkable exceptions with nodes of high degree
in large orbits
shown in Fig. \ref{dupgraphs} that cannot be found in the reference networks.
This is also well in accordance with the values found in Tab. \ref{duptab}. 
Different from the original network, in both reference networks the number of duplicate links is practically the same as the number of duplicate nodes.
\newline
From the Database of Interacting Proteins \cite{dip} and FlyBase \cite{flybase} we derived names and descriptions (if available) for the proteins in Fig. \ref{dupgraphs} as shown in Appendix B.
We find that duplicate proteins are likely to have similar functionality 
in accordance with results in the yeast PIN \cite{samanta:liang:2003}.
%%%%%%%%%%%%%%%%%%%%%%%%%%%%%%%%%%%%%%%%%%%%%%%%%%%%%%%%%%%%%%%%%%%%%%%%%%%%%
\section{Summary and conclusions}
Recent developments in the research on complex networks have brought up a better understanding of a network's topology and its connection to functionality. However, a comprehensive theory of networks incorporating classical graph theory as well as recent findings into a self-consistent framework has still to be worked out. Considering spectral graph theory to be a promising ansatz for this attempt, we have done a case study on the PIN of {\em Drosophila melanogaster}. The eigenvalues of a network's adjacency matrix (and of related matrices) provide information about a network's structural properties like the number of connected components, its diameter or characteristics of its degree distribution. Here, we have put special emphasis on the investigation of the discrete spectrum of a sparse network relating it to prevalent substructures. Although it will probably not be possible to derive the densities of specific subgraphs from the spectrum of a network we could show that structural prevalences on a more abstract level are reflected in the (discrete) spectrum of the PIN under investigation. While we here focused on the appearance of loops in subgraphs as well as the whole PIN future analysis might reveal further topological features.
\newline
Considering the evolutionary history of PINs we also discussed the appearance of proteins that share their neighbors together with the fingerprints of these structures that can be found in the network's spectrum. Studying structures of duplicate proteins in more details we find that they often have close functional relationships in accordance with earlier findings in yeast. 
\newline
The requirement applied here for the members of an orbit 
to show exactly the same
neighborhood is very restrictive, though required to allow for
transitivity. This might be generalized by the definition of a
similarity measure that quantifies the overlap of the neighborhoods
of two nodes. This similarity measure can be defined 
as a distance measure between nodes and the application of 
a clustering algorithm in the associated metric space might 
give further insight into local structures. 
\newline
This case study shows that a more systematic assessment of the relation between a network's spectral and topological properties has to be a topic of future research. It is a challenging task, however, it can bring important insight into a network's structure in a less biased and more systematic way than currently available.
\vspace*{0.3cm}
\newline
{\bf Acknowledgments:} C. Kamp would like to thank A. Bunten for providing computing facilities and software and for many helpful discussions on technical problems. I am as grateful to N. Farid, S.A. Teichmann and J. Leal for the discussions and many helpful comments on the manuscript. Also, we would like to acknowledge the hospitality of the department of physics of the university of Oslo during the period of finalizing this manuscript.  This work was supported by a fellowship within the Postdoc-Programme of the German Academic Exchange Service (DAAD).
\section*{Appendix A}
Let {$\bf A$} be a $N\times N$ matrix representing an undirected graph,
i.e. a symmetric matrix with entries $a_{ij}\in\{0,1\}$ and $a_{ii}=0$.
Let {$\bf D$ } be the matrix that is
obtained after $m$ perfect duplications of nodes or in other words
by $m$ duplications of rows and columns, respectively.
Let $i_1$,...,$i_l$, $l\in\{1,...,N\}$ the number (identifier)
of (mutually different) nodes that have
been duplicated and $m_{i_1}$,...,$m_{i_l}$ be the corresponding
number of duplications per node with
$m=\sum_{j=1}^l m_{i_j}$.
Let ${\bf A}_{i_j}$ the matrix ${\bf A}$ but with the element $a_{i_j i_j}$
replaced by $a_{i_j i_j}+\lambda$. Analogously, the matrix
${\bf A}_{i_1...i_l}$ corresponds to  the matrix ${\bf A}$ but
with $a_{i_1 i_1}$,...,$a_{i_l i_l}$ replaced by
$a_{i_1 i_1}+\lambda$,... $a_{i_l i_l}+\lambda$.
Let furthermore ${\bf I}$ be the identity matrix.
Then the following equation holds
\begin{eqnarray}\label{dupform}
&&\det({\bf D}-\lambda {\bf I})=\\ 
&& \!\!(-\lambda)^m \det({\bf A}-\lambda{\bf I})\nonumber\\
&+&\!\!(-\lambda)^m \sum_{r\leq l} m_{i_r} \det({\bf A}_{i_r}\!-\lambda {\bf I})\nonumber\\
&+&\!\!(-\lambda)^m\!\sum_{r\leq l-1}\sum_{r<j\leq l}
\!m_{i_r}m_{i_j}\det({\bf A}_{i_ri_j}\!-\lambda {\bf I})\nonumber\\
&+&\!\!(-\lambda)^m\!\sum_{r\leq l-2}\sum_{r<j\leq l-1}\sum_{j<s\leq
  l}\!\!\!m_{i_r}m_{i_j}m_{i_s}\det({\bf A}_{i_ri_ji_s}\!-\lambda {\bf I})\nonumber\\
&\vdots&\nonumber\\
&+&\!\!(-\lambda)^m\!\sum_{r\leq 1}\sum_{r<j\leq 2}...\sum_{x<y\leq
  l}\!m_{i_r}...m_{i_y}\det({\bf A}_{i_r...i_y}\!-\lambda{\bf I}).\nonumber
\end{eqnarray}
Note that the last term is equivalent to
$m_{i_1}...m_{i_l}\det({\bf A}_{i_1...i_l}-\lambda{\bf I})$.
It gets obvious from this formula that perfect duplication of nodes
only adds zeros to the spectrum of the graph.
\newline
Equation (\ref{dupform}) can be proven by induction. 
Considering a graph with adjacency matrix ${\bf A}$
in which an arbitrary node $i$ is duplicated 
$m$ times leading a duplication matrix ${\bf D}$ one can show that
\begin{equation}\label{step}
\det({\bf D}-\lambda {\bf I})=(-\lambda)^m[\det({\bf A}-\lambda {\bf I})
+m\det({\bf A}_{i}-\lambda {\bf I})].
\end{equation}
After validating the case of $l=0$, $m=0$ of equation (\ref{dupform})
we do the induction by evaluation of the
adjacency matrix $\tilde{{\bf D}}$
of a graph generated from the duplication graph represented by ${\bf D}$ 
by duplicating (a non-duplicate) node $i_{l+1}$  $m_{i_{l+1}}$ times.
Therefore, we  apply (\ref{step})
\begin{eqnarray*}
&&\det({\tilde{{\bf D}}}-\lambda {\bf I})\\
&=&
(-\lambda)^m_{i_{l+1}}[\det({\bf D}-\lambda {\bf I})
+m_{i_{l+1}}\det({\bf D}_{i_{l+1}}-\lambda {\bf I})]
\end{eqnarray*}
and derive $\det({\bf D}-\lambda {\bf I})$ and 
$\det({\bf D}_{i_{l+1}}-\lambda {\bf I})$ using the assumption 
(\ref{dupform}) yielding the formula (\ref{dupform}) for $m+m_{i_{l+1}}$
duplications of $l+1$ mutually different nodes.
\newline
As an example, this formula is applied to the $2\times2$-matrix ${\bf A}$
corresponding to two connected nodes. Then, $i_{1}=1$, ${i_2=2}$ and
one gets for the matrix $D$ after $m=m_1+m_2$ duplications:
\begin{eqnarray*}
\det({\bf D}-\lambda {\bf I})&=&(-\lambda)^m(\lambda^2-1-m_1-m_2-m_1m_2)\\
\lambda_{1;2}&=&\pm\sqrt{1+m+m_1m_2}.
\end{eqnarray*}
Translating this into the number of nodes per orbit $n_i=m_i+1$ leads
to the eigenvalues
\begin{eqnarray*}
\lambda_{1;2}=\pm\sqrt{n_1n_2}.
\end{eqnarray*}
%%%%%%%%%%%%%%%%%%%%%%%%%%%%%%%%%%%%%%%%%%%%%%%%%%%%%%%%%%%%%%%%%
\section*{Appendix B}
The following tables contain the information on the proteins shown in Fig. \ref{dupgraphs} extracted from the Database of Interacting Proteins \cite{dip} and FlyBase \cite{flybase}.
%\begin{table}[!h]
\begin{longtable}[!t]{ll}
DIP ID & Protein name/description\\
\hline
{\bf DIP:17489N} & CG11719-PA open reading frame, Mst98Ca,\\
& (Male-specific RNA 98Ca) \\
{\bf DIP:17490N} & CG18396-PA open reading frame, Mst98Cb,\\
&(Male-specific RNA 98Cb) \\
DIP:17144N &CG4015-PA open reading frame, Fcp3C,\\
& (Follicle cell protein 3C)\\
%& Polypeptides: Fcp3C-P1, Transcript: Fcp3C-RA\\
DIP:18312N &	CG17777-PA open reading frame\\
DIP:17051N &	CG17666-PA open reading frame\\
DIP:17123N &	CG15781-PA open reading frame\\
DIP:17121N & 	CG15032-PA open reading frame\\
DIP:17122N &	CG15489-PA open reading frame\\
DIP:20125N & 	CG1981-PA open reading frame, Thd1,\\
&G/T-mismatch-specific-thymine-DNA-\\
& glycosylase, double-stranded DNA-binding,\\
& mismatch repair\\
DIP:20084N &  	CG13363-PA open reading frame\\
DIP:19658N & CG12212-PA open reading frame, peb,\\
&(pebbled),\\
& transcription factor activity\\
%& (Polypeptides: peb-P1, Transcripts: peb-RA)\\
DIP:17485N &CG10154-PA open reading frame, \\
&structural constituent of peritrophic\\
& membrane, (sensu Insecta)\\
&\\
\caption{Proteins found in the 2-orbit with nodes of degree 10, 
both duplicates (bold) are male specific RNA (with corresponding polypeptides). }\label{tab210}
\end{longtable}	
%\end{table}
%%%%%%%%%%%%%%%%%%%%%%%%%%%%%%%%%%%%%%%%%%%%%%%%%%%%%%%%%%%%%%%%%%%%%%%%%%%%
\begin{longtable}[!h]{ll}
DIP ID & Protein name/description\\
\hline
{\bf DIP:18704N} & CG2789-PA open reading frame,\\
& bonzodiazepine receptor activity, \\
& transporter activity,\\
& metabolism and transport\\
{\bf DIP:18703N} & 	CG1341-PA open reading frame, Rpt1,\\
&endopeptidase activity, ATPase activity,\\
& proteolysis and peptidolysis\\
DIP:21261N &	CG3173-PA open reading frame\\
%&Domains: ARM repeat\\
DIP:20398N & 	CG12096-PA open reading frame\\
%&Domains: ARM repeat\\
DIP:17864N & CG10694-PA open reading frame\\
&damaged DNA-binding, base excision repair\\
%&Domains: UNA domain, ubiquities like\\
\hline
{\bf DIP:20457N} & CG12405-PA open reading frame, Prx2540-1,\\
&(Peroxiredoxin 2540),\\
&peroxidase, antioxidant activity,\\
& defense response, oxygen species metabolism\\
{\bf DIP:20458N} & CG12896-PA open reading frame,\\
&peroxidase activity\\
& defense response, oxygen species metabolism\\
DIP:20623N & CG9624-PA open reading frame\\
DIP:17811N & CG5576-PA open reading frame, imd,\\
&(immune deficiency),\\
&antimicrobial humoral response,\\
&(sensu Invertebrata)\\
DIP:17346N &CG12470-PA open reading frame\\
\hline
{\bf DIP:18389N} & 	CG18779-PA open reading frame\\
{\bf DIP:18387N} &  	open reading frame CG-10530/4-PA,\\
& Lcp65Ag1/Lcp65Ag2 protein,\\
& (larval cuticle protein),\\
& structural constituent of larval cuticle,\\
& (sensu Insecta),\\
& larval cuticle biosynthesis,\\
& (sensu Insecta)\\
DIP:18392N &	CG2082-PA open reading frame,\\
& signal transduction\\
DIP:18391N & 	CG16978-PA open reading frame\\
DIP:18390N & 	CG12907-PA open reading frame\\
&\\
\caption{Proteins found in a 2-orbit with nodes of degree 3, lines
separate different orbits, bold proteins are duplicates.}\label{tab23}
\end{longtable}
%\end{table}
%%%%%%%%%%%%%%%%%%%%%%%%%%%%%%%%%%%%%%%%%%%%%%%%%%%%%%%%%%%%%%%%%%%%%%%%%%%%%%%
%\begin{table}[!h]
\begin{longtable}[!h]{ll}
DIP ID & Protein name/description\\
\hline
{\bf DIP:20847N} & CG8284-PA open reading frame, UbcD4,\\
&(Ubiquitin conjugating enzyme 4),\\
& ubiquitin conjugating enzyme activity,\\
& ligase activity,\\
& protein metabolism, ubiquitin cycle\\
{\bf DIP:23198N} &	CG30344-PA open reading frame\\
{\bf DIP:18933N} &		CG10862-PA open reading frame,\\
& ubiquitin conjugating enzyme activity, \\
& ligase activity,\\
& protein metabolism\\
DIP:18935N & 	CG8974-PA open reading frame,\\
& transcription regulatory activity, \\
& ``nucleo-metabolism'', transcription,\\
DIP:18934N & 	CG32581-PA open reading frame,\\
& transcription regulatory activity, \\
& ``nucleo-metabolism'', transcription\\
\hline
{\bf DIP:20125N} & 	CG1981-PA open reading frame, Thd1,\\
&G/T-mismatch-specific-thymine-DNA-\\
& glycosylase, double-stranded DNA-binding,\\
& mismatch repair\\
{\bf DIP:19658N} & CG12212-PA open reading frame, peb,\\
&(pebbled),\\
& transcription factor activity\\
%& (Polypeptides: peb-P1, Transcripts: peb-RA)\\
{\bf DIP:20084N} &CG13363-PA open reading frame\\
DIP:17489N & CG11719-PA open reading frame, Mst98Ca,\\
& (Male-specific RNA 98Ca) \\
DIP:17490N & CG18396-PA open reading frame, Mst98Cb,\\
&(Male-specific RNA 98Cb) \\
&\\
\caption{{Proteins found in a 3-orbit with nodes of degree 2, lines separate different orbits, bold proteins are duplicates. Note, that Mst98Ca and Mst98Cb form a 2-orbit of degree 10, too (cf. Tab. \ref{tab210}).}}\label{tab32}
\end{longtable}
%\end{table}
%%%%%%%%%%%%%%%%%%%%%%%%%%%%%%%%%%%%%%%%%%%%%%%%%%%%%%%%%%%%%%%%%%%%%%%%%%%
%\begin{table}[!h]
\begin{longtable}[!h]{ll}
DIP ID & Protein name/description\\
\hline
{\bf DIP:19548N} &CG31366/18743-PA open reading frame,\\ 
&Hsp70A, (Heat shock protein 70A),\\
& heat, defense response, \\
&protein complex assembly and folding\\
{\bf DIP:19549N}  &CG31449/31359/6489-PA open reading frame,\\
& Hsp70B, (heat shock protein 70B),\\
& heat, defense response,\\
& protein complex assembly and folding\\
{\bf DIP:19551N} & open reading frame CG31449-PA, Hsp70Ba,\\
& (heat shock protein 70Ba),\\
& heat, defense response,\\
& protein complex assembly and folding\\
{\bf DIP:19552N} &	CG5834-PA open reading frame, Hsp70Bbb,\\
&(heat shock protein 70Bbb)\\
DIP:18493N & 	CG7945-PA open reading frame,\\
& chaperone activity\\
DIP:20272N &	CG5203-PA open reading frame, CHIP,\\
%&(Polypeptides: CHIP-P1, Transcripts: CHIP-RA)\\ 
&chaperone activity, protein folding \\
& and metabolism\\
DIP:20308N & CG32130-PA open reading frame\\
DIP:18578N & CG13165-PA open reading frame\\
%(DIP:19158N & Heat shock protein cognate 4, \\
%&Hsc70-4 (CG4264-PA))\\
%& chaperone activity, heat shock protein activity...\\
%& defense response, protein complex assembly,\\
%& protein folding...\\
%(DIP:23702N & CG7970-PA open reading frame)
&\\
\caption{Proteins found in the 4-orbit with nodes of degree 4,
 all duplicates (bold) are 
heat shock proteins (Hsp), released after heat shock or other stress.}\label{tab44}
\end{longtable}
%\end{table}
%%%%%%%%%%%%%%%%%%%%%%%%%%%%%%%%%%%%%%%%%%%%%%%%%%%%%%%%%%%%%%%%%%%%%%%%%%%%%
%\begin{table}[!h]
\begin{longtable}[!h]{ll}
DIP ID & Protein name/description\\
\hline
{\bf DIP:19748N} & 	CG1252-PA open reading frame, Ccp84Ab,\\
&(cuticle cluster 7),\\
& structural constituent of larval cuticle,\\
& (sensu Insecta)\\
{\bf DIP:19749N} & 	CG2360-PA open reading frame, Ccp84Aa,\\
&(cuticle cluster 8),\\
& structural constituent of larval cuticle,\\
&(sensu Insecta)\\
DIP:17392N & CG9949-PA open reading frame, sina,\\
& (seven in absentia),\\
& sensory organ development\\
%& glutathione transferase activity,\\
%& defense response, response to toxins\\
DIP:18536N & CG6615-PA open reading frame, scaf6,\\
& RNA binding, nuclear mRNA splicing\\
& via spliceosome, spliceosome complex\\
DIP:20225N & CG2341-PA open reading frame, Ccp84Ad,\\
&(cuticle cluster 5),\\
& structural constituent of larval cuticle,\\
& (sensu Insecta)\\
DIP:17713N & CG15422-PA open reading frame\\
DIP:17492N & CG12723-PA open reading frame\\
DIP:17076N & CG6945-PA open reading frame\\
DIP:17488N & CG11505-PB open reading frame\\
&\\
\caption{Proteins found in the 2-orbit with nodes of degree 7, duplicates (bold) are constituents of the larval cuticle. Note, that peb, Thd1, and CG13363-PA also form a 3-orbit with respect to Mst98Ca and Mst98Cb (cf. Tab. \ref{tab32}).}\label{tab27}
\end{longtable}	
%\end{table}

%\bibliographystyle{apsrev}
%\input{literatur_resub}
%\bibliography{/mn/tid/pgp-e1/ckamp/literature/lit}

\begin{thebibliography}{31}
\expandafter\ifx\csname natexlab\endcsname\relax\def\natexlab#1{#1}\fi
\expandafter\ifx\csname bibnamefont\endcsname\relax
  \def\bibnamefont#1{#1}\fi
\expandafter\ifx\csname bibfnamefont\endcsname\relax
  \def\bibfnamefont#1{#1}\fi
\expandafter\ifx\csname citenamefont\endcsname\relax
  \def\citenamefont#1{#1}\fi
\expandafter\ifx\csname url\endcsname\relax
  \def\url#1{\texttt{#1}}\fi
\expandafter\ifx\csname urlprefix\endcsname\relax\def\urlprefix{URL }\fi
\providecommand{\bibinfo}[2]{#2}
\providecommand{\eprint}[2][]{\url{#2}}

\bibitem[{\citenamefont{Strogatz}(2001)}]{strogatz:2001}
\bibinfo{author}{\bibfnamefont{S.~H.} \bibnamefont{Strogatz}},
  \bibinfo{journal}{Nature} \textbf{\bibinfo{volume}{410}},
  \bibinfo{pages}{268} (\bibinfo{year}{2001}).

\bibitem[{\citenamefont{Bollob{\'a}s}(1998)}]{bollobas:modernbook}
\bibinfo{author}{\bibfnamefont{B.}~\bibnamefont{Bollob{\'a}s}},
  \emph{\bibinfo{title}{Modern Graph Theory}} (\bibinfo{publisher}{Springer},
  \bibinfo{address}{New York}, \bibinfo{year}{1998}).

\bibitem[{\citenamefont{Bollob{\'a}s}(2001)}]{bollobas:randomgraphs}
\bibinfo{author}{\bibfnamefont{B.}~\bibnamefont{Bollob{\'a}s}},
  \emph{\bibinfo{title}{Random Graphs}} (\bibinfo{publisher}{Cambridge
  University Press}, \bibinfo{address}{Cambridge}, \bibinfo{year}{2001}).

\bibitem[{\citenamefont{Albert and
  Barab{´\'{a}}si}(2002)}]{albert:barabasi:2002}
\bibinfo{author}{\bibfnamefont{R.}~\bibnamefont{Albert}} \bibnamefont{and}
  \bibinfo{author}{\bibfnamefont{A.-L.} \bibnamefont{Barab{´\'{a}}si}},
  \bibinfo{journal}{Rev. Mod. Phys.} \textbf{\bibinfo{volume}{74}},
  \bibinfo{pages}{47} (\bibinfo{year}{2002}).

\bibitem[{\citenamefont{Chung}(1994)}]{chung:book}
\bibinfo{author}{\bibfnamefont{F.}~\bibnamefont{Chung}},
  \emph{\bibinfo{title}{Spectral Graph Theory}}, vol.~\bibinfo{volume}{92} of
  \emph{\bibinfo{series}{Regional Conference Series in Mathematics}}
  (\bibinfo{publisher}{American Mathematical Society Providence},
  \bibinfo{address}{Rhode Island}, \bibinfo{year}{1994}).

\bibitem[{\citenamefont{Cvetkovi{\'c} et~al.}(1980)\citenamefont{Cvetkovi{\'c},
  Doob, and Sachs}}]{cvetkovic:book}
\bibinfo{author}{\bibfnamefont{D.}~\bibnamefont{Cvetkovi{\'c}}},
  \bibinfo{author}{\bibfnamefont{M.}~\bibnamefont{Doob}}, \bibnamefont{and}
  \bibinfo{author}{\bibfnamefont{H.}~\bibnamefont{Sachs}},
  \emph{\bibinfo{title}{Spectra of graphs}}, Pure and applied mathematics
  (\bibinfo{publisher}{Academic Press}, \bibinfo{year}{1980}).

\bibitem[{\citenamefont{Cvetkovi{\'c} et~al.}(1988)\citenamefont{Cvetkovi{\'c},
  Doob, Gutman, and Torga{\v{s}}ev}}]{cvetkovic:annals}
\bibinfo{author}{\bibfnamefont{D.}~\bibnamefont{Cvetkovi{\'c}}},
  \bibinfo{author}{\bibfnamefont{M.}~\bibnamefont{Doob}},
  \bibinfo{author}{\bibfnamefont{I.}~\bibnamefont{Gutman}}, \bibnamefont{and}
  \bibinfo{author}{\bibfnamefont{A.}~\bibnamefont{Torga{\v{s}}ev}},
  \emph{\bibinfo{title}{Recent results in the theory of graph spectra}},
  no.~\bibinfo{number}{36} in \bibinfo{series}{Annals of discrete mathematics}
  (\bibinfo{publisher}{North Holland}, \bibinfo{year}{1988}).

\bibitem[{\citenamefont{Biggs}(1996)}]{biggs:book}
\bibinfo{author}{\bibfnamefont{N.}~\bibnamefont{Biggs}},
  \emph{\bibinfo{title}{Algebraic graph theory}} (\bibinfo{publisher}{Cambridge
  University Press}, \bibinfo{address}{Cambridge}, \bibinfo{year}{1996}).

\bibitem[{\citenamefont{Bauer and Golinelli}(2001)}]{bauer:golinelli:2001}
\bibinfo{author}{\bibfnamefont{M.}~\bibnamefont{Bauer}} \bibnamefont{and}
  \bibinfo{author}{\bibfnamefont{O.}~\bibnamefont{Golinelli}},
  \bibinfo{journal}{J. Stat. Phys.} \textbf{\bibinfo{volume}{103}},
  \bibinfo{pages}{301} (\bibinfo{year}{2001}).

\bibitem[{\citenamefont{Farkas et~al.}(2001)\citenamefont{Farkas, Der\'{e}nyi,
  Barab\'{a}si, and Vicsek}}]{farkas:vicsek:2001}
\bibinfo{author}{\bibfnamefont{I.}~\bibnamefont{Farkas}},
  \bibinfo{author}{\bibfnamefont{I.}~\bibnamefont{Der\'{e}nyi}},
  \bibinfo{author}{\bibfnamefont{A.-L.} \bibnamefont{Barab\'{a}si}},
  \bibnamefont{and} \bibinfo{author}{\bibfnamefont{T.}~\bibnamefont{Vicsek}},
  \bibinfo{journal}{Phys. Rev. E} \textbf{\bibinfo{volume}{64}},
  \bibinfo{pages}{026704} (\bibinfo{year}{2001}).

\bibitem[{\citenamefont{Dorogovtsev et~al.}(2003)\citenamefont{Dorogovtsev,
  Goltsev, Mendes, and Samukhin}}]{dorogovtsev:samukhin:2003}
\bibinfo{author}{\bibfnamefont{S.}~\bibnamefont{Dorogovtsev}},
  \bibinfo{author}{\bibfnamefont{A.}~\bibnamefont{Goltsev}},
  \bibinfo{author}{\bibfnamefont{J.}~\bibnamefont{Mendes}}, \bibnamefont{and}
  \bibinfo{author}{\bibfnamefont{A.}~\bibnamefont{Samukhin}},
  \bibinfo{journal}{Phys. Rev. E} \textbf{\bibinfo{volume}{68}},
  \bibinfo{pages}{046109} (\bibinfo{year}{2003}).
 % \urlprefix\url{http://arxiv.org/abs/cond-mat/0306340}.

\bibitem[{\citenamefont{Chung et~al.}(2003{\natexlab{a}})\citenamefont{Chung,
  Lu, and Vu}}]{chung:vu:2002}
\bibinfo{author}{\bibfnamefont{F.}~\bibnamefont{Chung}},
  \bibinfo{author}{\bibfnamefont{L.}~\bibnamefont{Lu}}, \bibnamefont{and}
  \bibinfo{author}{\bibfnamefont{V.}~\bibnamefont{Vu}},
  \bibinfo{journal}{Annals of Combinatorics} \textbf{\bibinfo{volume}{7}},
  \bibinfo{pages}{21} (\bibinfo{year}{2003}{\natexlab{a}}).
 % \bibinfo{note}{preprint from http://www.math.ucsd.edu/\~fan}.

\bibitem[{\citenamefont{Chung et~al.}(2003{\natexlab{b}})\citenamefont{Chung,
  Lu, and Vu}}]{chung:vu:2003}
\bibinfo{author}{\bibfnamefont{F.}~\bibnamefont{Chung}},
  \bibinfo{author}{\bibfnamefont{L.}~\bibnamefont{Lu}}, \bibnamefont{and}
  \bibinfo{author}{\bibfnamefont{V.}~\bibnamefont{Vu}}, \bibinfo{journal}{Proc.
  Natl. Acad. Sci. USA} \textbf{\bibinfo{volume}{100}}, \bibinfo{pages}{6313}
  (\bibinfo{year}{2003}{\natexlab{b}}).


\bibitem[{\citenamefont{Mihail and
  Papadimitriou}(2002)}]{mihail:papadimitriou:2002}
\bibinfo{author}{\bibfnamefont{M.}~\bibnamefont{Mihail}} \bibnamefont{and}
  \bibinfo{author}{\bibfnamefont{C.}~\bibnamefont{Papadimitriou}}, in
  \emph{\bibinfo{booktitle}{Proceedings of the 6th International Workshop on
  Randomization and Approximation Techniques}} (\bibinfo{year}{2002}), pp.
  \bibinfo{pages}{254--262}.

\bibitem[{\citenamefont{Goh et~al.}(2001)\citenamefont{Goh, Kahng, and
  Kim}}]{goh:kim:2001}
\bibinfo{author}{\bibfnamefont{K.-I.} \bibnamefont{Goh}},
  \bibinfo{author}{\bibfnamefont{B.}~\bibnamefont{Kahng}}, \bibnamefont{and}
  \bibinfo{author}{\bibfnamefont{D.}~\bibnamefont{Kim}},
  \bibinfo{journal}{Phys. Rev. E} \textbf{\bibinfo{volume}{64}},
  \bibinfo{pages}{051903} (\bibinfo{year}{2001}).

\bibitem[{\citenamefont{Golinelli}(2003)}]{golinelli:2003}
\bibinfo{author}{\bibfnamefont{O.}~\bibnamefont{Golinelli}},
  \bibinfo{journal}{http://arxiv.org/abs/cond-mat/0301437v1}
  (\bibinfo{year}{2003}).

\bibitem[{\citenamefont{Milo et~al.}(2002)\citenamefont{Milo, Shen-Orr,
  Itzkovitz, Kashtan, Chklovskii, and Alon}}]{milo:alon:2002}
\bibinfo{author}{\bibfnamefont{R.}~\bibnamefont{Milo}},
  \bibinfo{author}{\bibfnamefont{S.}~\bibnamefont{Shen-Orr}},
  \bibinfo{author}{\bibfnamefont{S.}~\bibnamefont{Itzkovitz}},
  \bibinfo{author}{\bibfnamefont{N.}~\bibnamefont{Kashtan}},
  \bibinfo{author}{\bibfnamefont{D.}~\bibnamefont{Chklovskii}},
  \bibnamefont{and} \bibinfo{author}{\bibfnamefont{U.}~\bibnamefont{Alon}},
  \bibinfo{journal}{Science} \textbf{\bibinfo{volume}{298}},
  \bibinfo{pages}{824} (\bibinfo{year}{2002}).

\bibitem[{\citenamefont{Milo et~al.}(2004)\citenamefont{Milo, Kashtan, Levitt,
  Shen-Orr, Ayzenshtat, Sheffer, and Alon}}]{milo:alon:2004}
\bibinfo{author}{\bibfnamefont{S.}~\bibnamefont{Milo},
  \bibfnamefont{R.~Itzkovitz}},
  \bibinfo{author}{\bibfnamefont{N.}~\bibnamefont{Kashtan}},
  \bibinfo{author}{\bibfnamefont{R.}~\bibnamefont{Levitt}},
  \bibinfo{author}{\bibfnamefont{S.}~\bibnamefont{Shen-Orr}},
  \bibinfo{author}{\bibfnamefont{I.}~\bibnamefont{Ayzenshtat}},
  \bibinfo{author}{\bibfnamefont{M.}~\bibnamefont{Sheffer}}, \bibnamefont{and}
  \bibinfo{author}{\bibfnamefont{U.}~\bibnamefont{Alon}},
  \bibinfo{journal}{Science} \textbf{\bibinfo{volume}{303}},
  \bibinfo{pages}{1538} (\bibinfo{year}{2004}).

\bibitem[{\citenamefont{Giot et~al.}(2003)\citenamefont{Giot, Bader, Brouwer,
  Chaudhuri, Kuang, Li, Hao, Ooi, Godwin, Vitols et~al.}}]{giot:rothberg:2003}
\bibinfo{author}{\bibfnamefont{L.}~\bibnamefont{Giot}},
  \bibinfo{author}{\bibfnamefont{J.}~\bibnamefont{Bader}},
  \bibinfo{author}{\bibfnamefont{C.}~\bibnamefont{Brouwer}},
  \bibinfo{author}{\bibfnamefont{A.}~\bibnamefont{Chaudhuri}},
  \bibinfo{author}{\bibfnamefont{B.}~\bibnamefont{Kuang}},
  \bibinfo{author}{\bibfnamefont{Y.}~\bibnamefont{Li}},
  \bibinfo{author}{\bibfnamefont{Y.}~\bibnamefont{Hao}},
  \bibinfo{author}{\bibfnamefont{C.}~\bibnamefont{Ooi}},
  \bibinfo{author}{\bibfnamefont{B.}~\bibnamefont{Godwin}},
  \bibinfo{author}{\bibfnamefont{E.}~\bibnamefont{Vitols}},
  \bibnamefont{et~al.}, \bibinfo{journal}{Science}
  \textbf{\bibinfo{volume}{302}}, \bibinfo{pages}{1727} (\bibinfo{year}{2003}).

\bibitem[{dip()}]{dip}
\emph{\bibinfo{title}{Database of Interacting Proteins ({DIP})}},
  \urlprefix\url{http://dip.doe-mbi.ucla.edu/}.

\bibitem[{\citenamefont{Maslov and Sneppen}(2002)}]{maslov:sneppen:2002}
\bibinfo{author}{\bibfnamefont{S.}~\bibnamefont{Maslov}} \bibnamefont{and}
  \bibinfo{author}{\bibfnamefont{K.}~\bibnamefont{Sneppen}},
  \bibinfo{journal}{Science} \textbf{\bibinfo{volume}{296}},
  \bibinfo{pages}{910} (\bibinfo{year}{2002}).

\bibitem[{\citenamefont{Milo et~al.}()\citenamefont{Milo, Shen-Orr, Itzkovitz,
  Kashtan, Chklovskii, and Alon}}]{mfinder}
\bibinfo{author}{\bibfnamefont{R.}~\bibnamefont{Milo}},
  \bibinfo{author}{\bibfnamefont{S.}~\bibnamefont{Shen-Orr}},
  \bibinfo{author}{\bibfnamefont{S.}~\bibnamefont{Itzkovitz}},
  \bibinfo{author}{\bibfnamefont{N.}~\bibnamefont{Kashtan}},
  \bibinfo{author}{\bibfnamefont{D.}~\bibnamefont{Chklovskii}},
  \bibnamefont{and} \bibinfo{author}{\bibfnamefont{U.}~\bibnamefont{Alon}},
  \emph{\bibinfo{title}{{\tt Mfinder} Software}},
  %\bibinfo{howpublished}{http://www.weizmann.ac.il/mcb/UriAlon/groupDownloadab%
%leData.html},
  \urlprefix\url{http://www.weizmann.ac.il/mcb/UriAlon/groupDownloadableData.html}.

\bibitem[{\citenamefont{Cvetkovi{\'c} et~al.}(1979)\citenamefont{Cvetkovi{\'c},
  Doob, and Sachs}}]{spectraofgraphs:book}
\bibinfo{author}{\bibfnamefont{D.}~\bibnamefont{Cvetkovi{\'c}}},
  \bibinfo{author}{\bibfnamefont{M.}~\bibnamefont{Doob}}, \bibnamefont{and}
  \bibinfo{author}{\bibfnamefont{H.}~\bibnamefont{Sachs}},
  \emph{\bibinfo{title}{Spectra of Graphs}}, Pure and Applied Mathematics
  (\bibinfo{publisher}{Academic Press}, \bibinfo{year}{1979}).

\bibitem[{\citenamefont{Ohno}(1970)}]{ohno:book}
\bibinfo{author}{\bibfnamefont{S.}~\bibnamefont{Ohno}},
  \emph{\bibinfo{title}{Evolution by gene duplication}}
  (\bibinfo{publisher}{Springer}, \bibinfo{address}{Berlin},
  \bibinfo{year}{1970}).

\bibitem[{\citenamefont{Teichmann and Babu}(2004)}]{teichmann:babu:2004}
\bibinfo{author}{\bibfnamefont{S.}~\bibnamefont{Teichmann}} \bibnamefont{and}
  \bibinfo{author}{\bibfnamefont{M.}~\bibnamefont{Babu}},
  \bibinfo{journal}{Nature Genetics} \textbf{\bibinfo{volume}{36}},
  \bibinfo{pages}{492} (\bibinfo{year}{2004}).

\bibitem[{\citenamefont{Amoutzias et~al.}(2004)\citenamefont{Amoutzias,
  Robertson, Oliver, and Bornberg-Bauer}}]{amoutzias:bauer:2004}
\bibinfo{author}{\bibfnamefont{G.}~\bibnamefont{Amoutzias}},
  \bibinfo{author}{\bibfnamefont{D.}~\bibnamefont{Robertson}},
  \bibinfo{author}{\bibfnamefont{S.}~\bibnamefont{Oliver}}, \bibnamefont{and}
  \bibinfo{author}{\bibfnamefont{E.}~\bibnamefont{Bornberg-Bauer}},
  \bibinfo{journal}{{EMBO}} \textbf{\bibinfo{volume}{5}}, \bibinfo{pages}{1}
  (\bibinfo{year}{2004}).

\bibitem[{\citenamefont{Kashtan et~al.}(2004)\citenamefont{Kashtan, Itzkovitz,
  Milo, and Alon}}]{kashtan:alon:2004}
\bibinfo{author}{\bibfnamefont{N.}~\bibnamefont{Kashtan}},
  \bibinfo{author}{\bibfnamefont{S.}~\bibnamefont{Itzkovitz}},
  \bibinfo{author}{\bibfnamefont{R.}~\bibnamefont{Milo}}, \bibnamefont{and}
  \bibinfo{author}{\bibfnamefont{U.}~\bibnamefont{Alon}},
  \bibinfo{journal}{Bioinformatics}  (\bibinfo{year}{2004}), \bibinfo{note}{in
  press}.

\bibitem[{fly()}]{flybase}
\emph{\bibinfo{title}{FlyBase, A Database of the Drosophila Genome}},
  \urlprefix\url{http://flybase.bio.indiana.edu/}.

\bibitem[{\citenamefont{Samanta and Liang}(2003)}]{samanta:liang:2003}
\bibinfo{author}{\bibfnamefont{M.}~\bibnamefont{Samanta}} \bibnamefont{and}
  \bibinfo{author}{\bibfnamefont{S.}~\bibnamefont{Liang}},
  \bibinfo{journal}{Proc. Natl. Acad. Sci. USA} \textbf{\bibinfo{volume}{100}},
  \bibinfo{pages}{12579} (\bibinfo{year}{2003}).

\bibitem[{\citenamefont{Kirkpatrick and
  Eggarter}(1972)}]{kirkpatrick:eggarter:1972}
\bibinfo{author}{\bibfnamefont{S.}~\bibnamefont{Kirkpatrick}} \bibnamefont{and}
  \bibinfo{author}{\bibfnamefont{T.}~\bibnamefont{Eggarter}},
  \bibinfo{journal}{Phys. Rev. B} \textbf{\bibinfo{volume}{6}},
  \bibinfo{pages}{3598} (\bibinfo{year}{1972}).

\bibitem[{\citenamefont{Evangelou}(1983)}]{evangelou:1983}
\bibinfo{author}{\bibfnamefont{S.}~\bibnamefont{Evangelou}},
  \bibinfo{journal}{Phys. Rev. B} \textbf{\bibinfo{volume}{27}},
  \bibinfo{pages}{1397} (\bibinfo{year}{1983}).

\end{thebibliography}
\end{document}